\documentstyle[preprint,pra,aps]{revtex}
\begin{document}
\draft
\title{Error correction code for protecting three-qubit quantum information against
erasures}
\author{Chui-Ping Yang \thanks{%
Email address: cpyang@floquet.chem.ku.edu} and Shih-I Chu \thanks{%
Email address: sichu@ku.edu}}
\address{Department of Chemistry, University of Kansas, and Kansas Center\\
for Advanced Scientific Computing, Lawrence, Kansas 66045}
\author{Siyuan Han \thanks{%
Email address: han@ku.edu}}
\address{Department of Physics and Astronomy, University of Kansas, Lawrence, Kansas\\
66045}
\maketitle

\begin{abstract}
We present a quantum error correction code which protects three quantum bits
(qubits) of quantum information against one erasure, i.e., a single-qubit
arbitrary error at a known position. To accomplish this, we encode the
original state by distributing quantum information over six qubits which is
the minimal number for the present task (see reference [1]). The encoding
and error recovery operations for such a code are presented. It is noted
that the present code is also a three-qubit quantum hidden information code
over each qubit. In addition, an encoding scheme for hiding $n$-qubit
quantum information over each qubit is proposed.
\end{abstract}

\date{\today }

\newpage

Quantum computing has become an active aspect of current research fields
with the discovery of Shor's algorithm for factorizing a large number [2-3].
It has become clear that quantum computer are in principle able to solve
hard computational problems more efficiently than present classical
computers [2-5]. However, the biggest difficulty inhibiting realizations is
the fragility of quantum states. Decoherence of qubits caused by the
interaction with environment will collapse the state of the quantum computer
and thus lead to the loss of information. To solve this problem, Shor, and
independently Stean, inspired by the theory of classical error correction,
proposed the first two quantum error correction codes (QECCs), i.e., the
nine-qubit code [6] and the seven-qubit code [7], which are able to correct
errors that occur during the store of qubits. Following this work, many new
QECCs have been discovered [8-21]. For the most general error model,
Laflamme et al. have shown that the smallest quantum error correction code,
for encoding one qubit of quantum information and correcting a single-qubit
arbitrary error at an unknown position, is the five-qubit code [8]. On the
other hand, apart from the QECCs, many alternative quantum codes have been
proposed, such as the quantum error preventing codes (based on the quantum
Zeno effect) [22-23] and the quantum error-avoiding codes (based on
decoherence-free subspaces (DFSs) [24-26]. Moreover, dynamical suppression
of decoherence [27-29] and noiseless subsystems [30-33] have been presented.

In 1997 M. Grassl et al. [34] considered an error model where the position
of the erroneous qubits is known. In accordance with classical coding
theory, they called this model the quantum erasure channel. Some physical
scenarios to determine the position of an error have been given [34]. In
their work, they showed that only four-qubit error correction code is
required to encode one qubit and correct one erasure ( i.e., a single-qubit
arbitrary error for which the position of the ``damaged'' qubit is known).
Also, they showed that two qubits of quantum information could be encoded
and one erasure could be corrected by extending such four-qubit code, in a
sense that only one additional qubit is required for encoding one
``message'' qubit on average. Clearly, this code is a very compact code for
protecting one or two qubits of quantum information as long as the position
of the ``bad'' qubit is known. In this paper we focus on how to protect
three qubits of quantum information against one erasure by a six-qubit code
described below. According to Ref [1], six qubits are the minimal number to
construct a code for the present purpose. The present code is also a
three-qubit quantum hidden information code over each qubit. In addition, we
propose an encoding scheme for hiding $n$-qubit quantum information over
each qubit, which provides a good illustration of the relationship between
quantum data hiding and QECC already noted by cleve et al. [35] and Cerf et
al. [36].

Protecting a few qubits of quantum information against decoherence is
important in quantum information and quantum computing. It is presumed that
the first prototype quantum computer will be small and quantum information
will be stored through only a few qubits. Moreover, there is much interest
arising from quantum computing network which is based on the connection of
locally distinct nodes each carrying out a small-scale quantum computing
[37]. In the following, we will first give a six-qubit code for protecting
three-qubit information against one erasure. We then discuss how to perform
the encoding, decoding and error recovery operations.

The Hilbert space of a three-qubit system is a tensor product of
two-dimensional spaces $C_2$ (qubits), i.e., $C=C_2^{\otimes 3}.$ An
arbitrary state of three qubits (labeled by 1, 2 and 3) can be expanded as
follows 
\begin{equation}
\left| \psi \right\rangle _{123}=\alpha _0\left| 000\right\rangle +\alpha
_1\left| 001\right\rangle +\alpha _2\left| 010\right\rangle +\alpha _3\left|
011\right\rangle +\alpha _4\left| 100\right\rangle +\alpha _5\left|
101\right\rangle +\alpha _6\left| 110\right\rangle +\alpha _7\left|
111\right\rangle ,
\end{equation}
where $\sum\limits_{i=0}^7\left| \alpha _i\right| ^2=1;$ $\left\{ \left|
ijk\right\rangle \right\} $ forms a set of complete orthogonal states in the
eight-dimensional space, $i,j,k\in \left\{ 0,1\right\} ;$ and we are taking
the $\left| 0\right\rangle $ \ and $\left| 1\right\rangle $ states of a
qubit to correspond to the `` down'' and ``up'' states, respectively, of a
fictitious spin $\frac 12$ particle. Using three ancillary qubits ($%
1^{\prime },$ $2^{\prime },$ $3^{\prime }$), we encode the original state
into 
\begin{equation}
\left| \psi \right\rangle _L=\alpha _0\left| 0\right\rangle _L+\alpha
_1\left| 1\right\rangle _L+\alpha _2\left| 2\right\rangle _L+\alpha _3\left|
3\right\rangle _L+\alpha _4\left| 4\right\rangle _L+\alpha _5\left|
5\right\rangle _L+\alpha _6\left| 6\right\rangle _L+\alpha _7\left|
7\right\rangle _L,
\end{equation}
where the eight logical states are 
\begin{eqnarray}
\left| 0\right\rangle _L &=&\left( \left| 000\right\rangle +\left|
111\right\rangle \right) \otimes \left( \left| 000\right\rangle +\left|
111\right\rangle \right) ,  \nonumber \\
\left| 1\right\rangle _L &=&\left( \left| 000\right\rangle -\left|
111\right\rangle \right) \otimes \left( \left| 000\right\rangle -\left|
111\right\rangle \right) ,  \nonumber \\
\left| 2\right\rangle _L &=&\left( \left| 010\right\rangle +\left|
101\right\rangle \right) \otimes \left( \left| 010\right\rangle +\left|
101\right\rangle \right) ,  \nonumber \\
\left| 3\right\rangle _L &=&\left( \left| 010\right\rangle -\left|
101\right\rangle \right) \otimes \left( \left| 010\right\rangle -\left|
101\right\rangle \right) ,  \nonumber \\
\left| 4\right\rangle _L &=&\left( \left| 100\right\rangle +\left|
011\right\rangle \right) \otimes \left( \left| 100\right\rangle +\left|
011\right\rangle \right) ,  \nonumber \\
\left| 5\right\rangle _L &=&\left( \left| 100\right\rangle -\left|
011\right\rangle \right) \otimes \left( \left| 100\right\rangle -\left|
011\right\rangle \right) ,  \nonumber \\
\left| 6\right\rangle _L &=&\left( \left| 110\right\rangle +\left|
001\right\rangle \right) \otimes \left( \left| 110\right\rangle +\left|
001\right\rangle \right) ,  \nonumber \\
\left| 7\right\rangle _L &=&\left( \left| 110\right\rangle -\left|
001\right\rangle \right) \otimes \left( \left| 110\right\rangle -\left|
001\right\rangle \right)
\end{eqnarray}
(here, for every logical state, the left part of the product corresponds to
the three ``message'' qubits while the right part of the product corresponds
to the three ancillary qubits, and the arrangement sequence of the six
qubits is 1, 2, 3, $1^{\prime },$ $2^{\prime }$ and $3^{\prime }$ from left
to right; to simplify the notation, normalization factors are omitted here
and in the remainder of this section).

Let us first briefly review some basics of quantum error correction codes.
It has been shown that one can model the errors by the use of error
operators $A.$ For the general case, Kill and Laflamme [18] derived the
following necessary and sufficient conditions on quantum error correction
codes 
\begin{equation}
\left\langle i_L\right| A_a^{+}A_b\left| i_L\right\rangle =\left\langle
j_L\right| A_a^{+}A_b\left| j_L\right\rangle ,
\end{equation}
and 
\begin{equation}
\left\langle i_L\right| A_a^{+}A_b\left| j_L\right\rangle =0\text{ \quad for 
}\left\langle i_L\right| \left. j_L\right\rangle =0,
\end{equation}
where $\left| i_L\right\rangle $ and $\left| j_L\right\rangle $ are any two
orthonormal basis states of the code (i.e., any two logical states). For the
purpose of error correction, it is enough to consider errors of the type $%
\sigma _x$ (bit flip), $\sigma _z$ (phase flip), and $\sigma _y$ (bit and
phase flip), since, by linearity, a code that can correct these errors can
correct any arbitrary errors [9]. For a [$n,k,t$] code, i.e., a code
encoding $k$ qubits through $n$ qubits and correcting $t$ errors at most,
the error operators $\left\{ A_a\right\} $ are the tensor product of the
identity on $n-t$ qubits and $t$ one-bit error operators on the altered
qubits. The one-bit error operators are any linear combinations of the
algebra basis $\left\{ 1,\sigma _x,\sigma _{y,}\sigma _z\right\} .$

The above conditions have been generalized to the quantum erasure channel
[34, 36]. Since the positions of the errors are known, it is not necessary
to separate the spaces which correspond to errors at different positions.
For the case of correcting erasure errors, the error operators $A_a$ and $%
A_b $ differ from each other by one-bit error operators at the same
positions only. Since the product of such $t$-error operators is also a $t$%
-error operator which can be written as a linear combination of the $A_a,$
it follows from Eqs. (4) and (5) that the necessary and sufficient
conditions corresponding to the erasure-correcting case will be [34, 36] 
\begin{eqnarray}
\left\langle i_L\right| A_a\left| i_L\right\rangle &=&\left\langle
j_L\right| A_a\left| j_L\right\rangle , \\
\left\langle i_L\right| A_a\left| j_L\right\rangle &=&0\text{ \quad for }%
\left\langle i_L\right| \left. j_L\right\rangle =0.
\end{eqnarray}

Now we give the interpretations of the encoding (3) in terms of error
correction codes. For the case of one erasure, the error operators $A_a$ in
Eqs. (6) and (7) are the one-bit error operators for the ``bad'' qubit,
which are any linear combinations of the algebra basis $\left\{ 1,\sigma
_x,\sigma _{y,}\sigma _z\right\} .$ One can easily verify that no matter
which qubit goes ``bad'', any two of the eight logical states (3) satisfy
the above conditions (6) and (7). Thus, these logical states in (3) can be
regarded as an erasure-correcting code: it can, in principle, encode three
qubits and correct one erasure. In the following, we will show explicitly
how this can be done.

The encoding (3) can be fulfilled by the quantum CNOT (controlled-NOT)
operations $C_{ij}$, where the first subscript of $C_{ij}$ refers to the
control bit and the second to the target. The three ancillary qubits 1$%
^{\prime },$ $2^{\prime }$ and 3$^{\prime }$ are initially in the state $%
\left| 000\right\rangle $. Throughout this paper, every joint operation will
follow the sequence from right to left. Let a joint encoding operation on
the six qubits 
\begin{equation}
U_e=C_{3^{\prime }2^{\prime }}C_{3^{\prime }1^{\prime }\text{ }%
}C_{32}C_{31}H_{3^{\prime }}H_3C_{33^{\prime }}C_{22^{\prime }}C_{11^{\prime
}},
\end{equation}
where $H_i$ is a Hadamard transformation on the qubit $i$ which sends $%
\left| 0\right\rangle \rightarrow \left( \left| 0\right\rangle +\left|
1\right\rangle \right) $ and $\left| 1\right\rangle \rightarrow \left(
\left| 0\right\rangle -\left| 1\right\rangle \right) ,$ thus we have 
\begin{equation}
U_e\left( \left| \psi \right\rangle _{123}\left| 000\right\rangle
_{1^{\prime }2^{\prime }3^{\prime }}\right) =\text{ }\left| \psi
\right\rangle _L.
\end{equation}

One can certainly envision situations where one might, in fact, know where
the error has occurred (by using the methods for determining the position of
an error [34]). Let us first consider the case in which qubit 1 undergoes
decoherence. Because $\left| 0\right\rangle $ and $\left| 1\right\rangle $
form a basis for the qubit 1, we need only know what happens to these two
states. In general, the decoherence process must be 
\begin{eqnarray}
\left| e_0\right\rangle \left| 0\right\rangle &\rightarrow &\left| \epsilon
_0\right\rangle \left| 0\right\rangle +\left| \epsilon _1\right\rangle
\left| 1\right\rangle ,  \nonumber \\
\left| e_0\right\rangle \left| 1\right\rangle &\rightarrow &\left| \epsilon
_0^{\prime }\right\rangle \left| 0\right\rangle +\left| \epsilon _1^{\prime
}\right\rangle \left| 1\right\rangle ,
\end{eqnarray}
where $\left| \epsilon _0\right\rangle ,\left| \epsilon _1\right\rangle
,\left| \epsilon _0^{\prime }\right\rangle $ and $\left| \epsilon _1^{\prime
}\right\rangle $ are appropriate environment states, not necessarily
orthogonal or normalized and $\left| e_0\right\rangle $ is the initial state
of the environment. As will be shown below, during the restoration operation
there is no need of performing any operations on the qubit 1. For the
simplicity, we can rewrite Eq. (10) as 
\begin{eqnarray}
\left| e_0\right\rangle \left| 0\right\rangle &\rightarrow &\left| 
\widetilde{0}\right\rangle ,  \nonumber \\
\left| e_0\right\rangle \left| 0\right\rangle &\rightarrow &\left| 
\widetilde{1}\right\rangle ,
\end{eqnarray}
where the above environment states $\left| \epsilon _0\right\rangle ,$ $%
\left| \epsilon _1\right\rangle ,$ $\left| \epsilon _0^{\prime
}\right\rangle $ and $\left| \epsilon _1^{\prime }\right\rangle $ have been
included in $\left| \widetilde{0}\right\rangle $ and $\left| \widetilde{1}%
\right\rangle .$ Let us now see what will happen to the encoded state $%
\left| \psi \right\rangle _L.$ After decoherence, it goes to 
\begin{equation}
\left| \psi \right\rangle _L\otimes \left| e_0\right\rangle =\alpha _0\left| 
\widetilde{0}\right\rangle _L+\alpha _1\left| \widetilde{1}\right\rangle
_L+\alpha _2\left| \widetilde{2}\right\rangle _L+\alpha _3\left| \widetilde{3%
}\right\rangle _L+\alpha _4\left| \widetilde{4}\right\rangle _L+\alpha
_5\left| \widetilde{5}\right\rangle _L+\alpha _6\left| \widetilde{6}%
\right\rangle _L+\alpha _7\left| \widetilde{7}\right\rangle _L,
\end{equation}
where 
\begin{eqnarray}
\left| \widetilde{0}\right\rangle _L &=&\left( \left| \widetilde{0}%
00\right\rangle +\left| \widetilde{1}11\right\rangle \right) \otimes \left(
\left| 000\right\rangle +\left| 111\right\rangle \right) ,  \nonumber \\
\left| \widetilde{1}\right\rangle _L &=&\left( \left| \widetilde{0}%
00\right\rangle -\left| \widetilde{1}11\right\rangle \right) \otimes \left(
\left| 000\right\rangle -\left| 111\right\rangle \right) ,  \nonumber \\
\left| \widetilde{2}\right\rangle _L &=&\left( \left| \widetilde{0}%
10\right\rangle +\left| \widetilde{1}01\right\rangle \right) \otimes \left(
\left| 010\right\rangle +\left| 101\right\rangle \right) ,  \nonumber \\
\left| \widetilde{3}\right\rangle _L &=&\left( \left| \widetilde{0}%
10\right\rangle -\left| \widetilde{1}01\right\rangle \right) \otimes \left(
\left| 010\right\rangle -\left| 101\right\rangle \right) ,  \nonumber \\
\left| \widetilde{4}\right\rangle _L &=&\left( \left| \widetilde{1}%
00\right\rangle +\left| \widetilde{0}11\right\rangle \right) \otimes \left(
\left| 100\right\rangle +\left| 011\right\rangle \right) ,  \nonumber \\
\left| \widetilde{5}\right\rangle _L &=&\left( \left| \widetilde{1}%
00\right\rangle -\left| \widetilde{0}11\right\rangle \right) \otimes \left(
\left| 100\right\rangle -\left| 011\right\rangle \right) ,  \nonumber \\
\left| \widetilde{6}\right\rangle _L &=&\left( \left| \widetilde{1}%
10\right\rangle +\left| \widetilde{0}01\right\rangle \right) \otimes \left(
\left| 110\right\rangle +\left| 001\right\rangle \right) ,  \nonumber \\
\left| \widetilde{7}\right\rangle _L &=&\left( \left| \widetilde{1}%
10\right\rangle -\left| \widetilde{0}01\right\rangle \right) \otimes \left(
\left| 110\right\rangle -\left| 001\right\rangle \right) .
\end{eqnarray}

Comparing Eq. (13 ) with Eq. (3), one can see that for each ``bad'' logical
state in (13), the right part of the product, which corresponds to the
encoding of the three ancillary qubits, is intact. We can first perform a
unitary transformation on the three ancillary qubits which we regard as the
partial decoding operation (since the qubits 1, 2 and 3 are not involved in
the decoding operation). The decoding operation is shown as follows 
\begin{equation}
U_d=H_{3^{\prime }}C_{3^{\prime }2^{\prime }}C_{3^{\prime }1^{\prime }}.
\end{equation}
After decoding, we have 
\begin{eqnarray}
\left| \widetilde{0}\right\rangle _L &\rightarrow &\left( \left| \widetilde{0%
}00\right\rangle +\left| \widetilde{1}11\right\rangle \right) \otimes \left|
000\right\rangle ,  \nonumber \\
\left| \widetilde{1}\right\rangle _L &\rightarrow &\left( \left| \widetilde{0%
}00\right\rangle -\left| \widetilde{1}11\right\rangle \right) \otimes \left|
001\right\rangle ,  \nonumber \\
\left| \widetilde{2}\right\rangle _L &\rightarrow &\left( \left| \widetilde{0%
}10\right\rangle +\left| \widetilde{1}01\right\rangle \right) \otimes \left|
010\right\rangle ,  \nonumber \\
\left| \widetilde{3}\right\rangle _L &\rightarrow &\left( \left| \widetilde{0%
}10\right\rangle -\left| \widetilde{1}01\right\rangle \right) \otimes \left|
011\right\rangle ,  \nonumber \\
\left| \widetilde{4}\right\rangle _L &\rightarrow &\left( \left| \widetilde{1%
}00\right\rangle +\left| \widetilde{0}11\right\rangle \right) \otimes \left|
100\right\rangle ,  \nonumber \\
\left| \widetilde{5}\right\rangle _L &\rightarrow &\left( \left| \widetilde{1%
}00\right\rangle -\left| \widetilde{0}11\right\rangle \right) \otimes \left|
101\right\rangle ,  \nonumber \\
\left| \widetilde{6}\right\rangle _L &\rightarrow &\left( \left| \widetilde{1%
}10\right\rangle +\left| \widetilde{0}01\right\rangle \right) \otimes \left|
110\right\rangle ,  \nonumber \\
\left| \widetilde{7}\right\rangle _L &\rightarrow &\left( \left| \widetilde{1%
}10\right\rangle -\left| \widetilde{0}01\right\rangle \right) \otimes \left|
111\right\rangle .
\end{eqnarray}

What we need to do now is to perform an error recovery operation in order to
extract the original state (1). It can be done by a unitary transformation
on the qubits 2, 3, 1$^{\prime },2^{\prime }$ and $3^{\prime },$ which is
described by 
\begin{equation}
U_r=T_{1^{\prime }3^{\prime }2}Z_{3^{\prime }2}T_{1^{\prime }3^{\prime
}2}C_{2^{\prime }2}C_{1^{\prime }2}C_{1^{\prime }3},
\end{equation}
where $T_{1^{\prime }3^{\prime }2}$ is a Toffoli gate operation [38], and $%
Z_{3^{\prime }2}$ is a controlled Pauli $\sigma _z$ operation. A Toffoli
gate operation $T_{ijk}$ has the two control bits corresponding to the first
two subscripts ($i$, $j$), and the target bit $k.$ When the two control bits
are in the state $\left| 11\right\rangle $, the state of the target bit will
change, following $\left| 0\right\rangle \rightarrow \left| 1\right\rangle $
and $\left| 1\right\rangle \rightarrow \left| 0\right\rangle ;$ while when
the two control bits are in the state $\left| 00\right\rangle ,\left|
01\right\rangle $ or $\left| 10\right\rangle ,$ the state of the target bit
will be invariant. A controlled Pauli $\sigma _z$ operation $Z_{ij}$ has the
control bit $i$ and the target bit $j$, which sends the state of the target
bit $\left| 0\right\rangle \rightarrow \left| 0\right\rangle $ and $\left|
1\right\rangle \rightarrow -\left| 1\right\rangle $ when the control bit is
in the state $\left| 1\right\rangle $; otherwise, when the control bit is in 
$\left| 0\right\rangle ,$ the state of the target bit will not change. One
can easily verify that after the operation $U_r$, the system composed of the
six qubits and the environment will be in the state 
\begin{equation}
\left( \left| \widetilde{0}00\right\rangle +\left| \widetilde{1}%
11\right\rangle \right) \otimes \left| \psi \right\rangle _{1^{\prime
}2^{\prime }3^{\prime }},
\end{equation}
where 
\begin{eqnarray}
\left| \psi \right\rangle _{1^{\prime }2^{\prime }3^{\prime }} &=&\alpha
_0\left| 000\right\rangle +\alpha _1\left| 001\right\rangle +\alpha _2\left|
010\right\rangle +\alpha _3\left| 011\right\rangle  \nonumber \\
&&\ +\alpha _4\left| 100\right\rangle +\alpha _5\left| 101\right\rangle
+\alpha _6\left| 110\right\rangle +\alpha _7\left| 111\right\rangle .
\end{eqnarray}

From Eqs. (17-18), one can see that the above restoration operation is
actually a disentangling operation, which has made the three qubits $%
1^{\prime },2^{\prime }$ and $3^{\prime }$ no longer entangled with the
remaining system (i.e., the three qubits 1, 2, 3 and the environment). Even
though the three qubits 1, 2 and 3 are entangled with the environment, the
information, originally carried by the qubits 1, 2 and 3, has been
completely transferred into the three qubits 1$^{\prime },2^{\prime }$ and 3$%
^{\prime },$ and the original state (1) has been exactly reconstructed
through the three qubits 1$^{\prime },2^{\prime }$ and 3$^{\prime }.$

It is straightforward to extract the original state when the error occurs on
the qubit 2 or 3$.$ To simplify our presentation, however, we will not give
a detailed discussion. In the case of qubit 2 or qubit 3 going ``bad'', the
decoding operation is the same as above. If the qubit 2 goes ``bad'', the
error recovery operation will be $T_{2^{\prime }3^{\prime }1}Z_{3^{\prime
}1}T_{2^{\prime }3^{\prime }1}C_{1^{\prime }1}C_{2^{\prime }1}C_{2^{\prime
}3};$ while when the qubit 3 goes ``bad'', the error recovery operation is
much simpler, i.e., $Z_{3^{\prime }2}C_{2^{\prime }2}C_{1^{\prime }1}$.
After performing the error recovery operations, the final state,
corresponding to the case when the error occurs on the qubit 2 or 3, will be 
\begin{equation}
\left( \left| 0\widetilde{0}0\right\rangle +\left| 1\widetilde{1}%
1\right\rangle \right) \otimes \left| \psi \right\rangle _{1^{\prime
}2^{\prime }3^{\prime }},
\end{equation}
or 
\begin{equation}
\left( \left| 00\widetilde{0}\right\rangle +\left| 11\widetilde{1}%
\right\rangle \right) \otimes \left| \psi \right\rangle _{1^{\prime
}2^{\prime }3^{\prime }}.
\end{equation}

In above we discussed how to recover the original state when the qubit 1, 2
or 3 undergoes decoherence. From Eq. (3) one can easily see that for each
logical state, the qubits 1, 2, 3 and the qubits $1^{\prime },$ $2^{\prime
}, $ $3^{\prime }$ are in the same GHZ states, i.e., each logical state is a
product of two copies of a three-qubit GHZ state. Thus, the decoding and
error recovery operations for the case of the qubit $1^{\prime },$ $%
2^{\prime }$ or 3$^{\prime }$ going ``bad'' are similar to those,
respectively, for the case of the qubit 1, 2 or 3 going ``bad''. The only
thing to be noted is that when the qubits $1^{\prime },$ $2^{\prime }$ or 3$%
^{\prime }$ goes ``bad'', the subscripts (1$^{\prime }$, 2$^{\prime },$ $%
3^{\prime },$ 1, 2, 3), which are involved in the above decoding and
error-recovery unitary transformations, need to be permuted into (1, 2$,$ $%
3, $ $1^{\prime },$ $2^{\prime }$, 3$^{\prime })$, respectively. Thus, we
have (a) when the qubits 1$^{\prime },$ $2^{\prime }$ or $3^{\prime }$ goes
``bad'', the decoding operation is given by $H_3C_{32}C_{31};$ (b) for the
case of the qubit $1^{\prime },$ $2^{\prime }$ or $3^{\prime }$ going
``bad'', the error recovery operation is given by $T_{132^{\prime
}}Z_{32^{\prime }}T_{132^{\prime }}C_{22^{\prime }}C_{12^{\prime
}}C_{13^{\prime }},$ $T_{231^{\prime }}Z_{31^{\prime }}T_{231^{\prime
}}C_{11^{\prime }}C_{21^{\prime }}C_{23^{\prime }}$ or $Z_{32^{\prime
}}C_{22^{\prime }}C_{11^{\prime }},$ respectively. After performing the
decoding and error recovery operations, the original state will be restored
through the qubits 1, 2 and 3; while the qubits 1$^{\prime },2^{\prime }$
and $3^{\prime }$ are entangled with the environment.

It should be mentioned that the above decoherence process (10), in fact,
corresponds to the case when qubits are represented by ideal ``two-state''
or ``two-level'' systems. In most cases, physical systems (particles or
solid state devices) may have many levels, such as atoms, ions and SQUIDs.
If a qubit is represented by a two-dimensional (2D) subspace of the Hilbert
space of a multi-level physical system, the interaction with environment may
lead to the leakage of a qubit out of the 2D subspace (i.e., the space
spanned by the two states $\left| 0\right\rangle $ and $\left|
1\right\rangle $ of a qubit). The decoherence process, therefore, is given by

\begin{eqnarray}
\left| e_0\right\rangle \left| 0\right\rangle &\rightarrow &\left| \epsilon
_0\right\rangle \left| 0\right\rangle +\left| \epsilon _1\right\rangle
\left| 1\right\rangle +\sum\limits_{i\neq 0,1}\left| \epsilon
_i\right\rangle \left| i\right\rangle ,  \nonumber \\
\left| e_0\right\rangle \left| 1\right\rangle &\rightarrow &\left| \epsilon
_0^{\prime }\right\rangle \left| 0\right\rangle +\left| \epsilon _1^{\prime
}\right\rangle \left| 1\right\rangle +\sum\limits_{i\neq 0,1}\left| \epsilon
_i^{\prime }\right\rangle \left| i\right\rangle ,
\end{eqnarray}
where $\left\{ \left| i\right\rangle \right\} $, together with $\left|
0\right\rangle $ and $\left| 1\right\rangle ,$ forms a complete orthogonal
basis of a multi-level system, and $\left| \epsilon _i\right\rangle $, $%
\left| \epsilon _i^{\prime }\right\rangle $ are environment states. Note
that during the above restoration operation, there is no need of performing
any operations on the ``bad''qubit. Thus, for the case when a qubit is
represented by a 2D subspace of a multi-level physical system and
decoherence happens like (21), one can still protect an arbitrary state of
three qubits against one erasure by using the code and following the
restoration operations described above.

It is noted that for some special types of three-qubit state, it is possible
that the protection against one erasure may be done by a code with a smaller
number of qubits. For example, one can show that the following three-qubit
states 
\begin{equation}
\alpha \left| 001\right\rangle +\beta \left| 010\right\rangle +\gamma \left|
100\right\rangle
\end{equation}
(which, in the case of $\left| \alpha \right| =\left| \beta \right| =\left|
\gamma \right| =\frac 1{\sqrt{2}},$ are called ``entangled W states'' [39]
that have attracted much interest recently) can be protected against one
erasure through the following five-qubit code 
\begin{eqnarray}
\left| 001\right\rangle &\rightarrow &\left| 00001\right\rangle +\left|
11110\right\rangle ,  \nonumber \\
\left| 010\right\rangle &\rightarrow &\left| 00100\right\rangle +\left|
11011\right\rangle ,  \nonumber \\
\left| 100\right\rangle &\rightarrow &\left| 00010\right\rangle +\left|
11101\right\rangle .
\end{eqnarray}

In above, it has been shown that the code (3) can be used to protect three
qubits of quantum information against one erasure. According to the theory
about ``connection between quantum information hiding and QECC'' [35,36],
this code should be also a quantum code for hiding three qubits of quantum
information over each qubit. This can be easily understood, since the
``bad'' qubit is not involved in the above restoration operation (i.e., it
does not contain any information so that it can be ``thrown away'' without
affecting the recovery of the original message). In the remainder of this
paper, we will give an encoding scheme for hide $n$-qubit quantum
information over each qubit.

An arbitrary state of $n$ ``message'' qubits can be written as follows 
\begin{equation}
\left| \psi \right\rangle =\sum_{i=0}^{2^n}\alpha _i\left| i\right\rangle ,
\end{equation}
where $\sum\limits_{i=0}^{2^n}\left| \alpha _i\right| ^2=1$; and $\left|
i\right\rangle $ represents a general basis state of $n$ qubits with the
integer $i$ corresponding to its binary decomposition. To hide $n$-qubit
quantum information, we can use $n$ ancillary qubits to encode the state
(24) into 
\begin{equation}
\left| \psi \right\rangle _L=\sum_{i=0}^{2^n}\alpha _i\left| \psi ^{\left(
i\right) }\right\rangle _{12...n}\otimes \left| \psi ^{\left( i\right)
}\right\rangle _{1^{\prime }2^{\prime }...n^{\prime }},\qquad
\end{equation}
where $\left| \psi _{12...n}^{\left( i\right) }\right\rangle $ and $\left|
\psi ^{\left( i\right) }\right\rangle _{1^{\prime }2^{\prime }...n^{\prime
}} $ are the two $n$-qubit GHZ states, respectively, corresponding to the $n$
``message'' qubits (1, 2, $\cdot \cdot \cdot ,$ $n$) and the $n$ ancillary
qubits ( $1^{\prime },2^{\prime },\cdot \cdot \cdot ,n^{\prime }$), which
are given by 
\begin{eqnarray}
\left| \psi ^{\left( i\right) }\right\rangle _{12...n} &=&\frac 1{\sqrt{2}}%
\left[ \left| u_1^{\left( i\right) }u_2^{\left( i\right) }\cdot \cdot \cdot
u_n^{\left( i\right) }\right\rangle \pm \left| \overline{u}_1^{\left(
i\right) }\overline{u}_2^{\left( i\right) }\cdot \cdot \cdot \overline{u}%
_n^{\left( i\right) }\right\rangle \right] ,  \nonumber \\
\left| \psi ^{\left( i\right) }\right\rangle _{1^{\prime }2^{\prime
}...n^{\prime }} &=&\frac 1{\sqrt{2}}\left[ \left| v_{1^{\prime }}^{\left(
i\right) }v_{2^{\prime }}^{\left( i\right) }\cdot \cdot \cdot v_{n^{\prime
}}^{\left( i\right) }\right\rangle \pm \left| \overline{v}_{1^{\prime
}}^{\left( i\right) }\overline{v}_{2^{\prime }}^{\left( i\right) }\cdot
\cdot \cdot \overline{v}_{n^{\prime }}^{\left( i\right) }\right\rangle
\right]
\end{eqnarray}
(here, $\left| u_k^{\left( i\right) }\right\rangle $ and $\left| \overline{u}%
_k^{\left( i\right) }\right\rangle $ represent two orthogonal states of the
``message'' qubit $k$, $\overline{u}_k^{\left( i\right) }$ $=1-u_k^{\left(
i\right) }$ and $u_k^{\left( i\right) }\in \{0,1\}$; the same notation holds
for the two orthogonal states $\left| v_{k^{\prime }}^{\left( i\right)
}\right\rangle $ and $\left| \overline{v}_{k^{\prime }}^{\left( i\right)
}\right\rangle $ of the ancillary qubit $k^{\prime }$).

Since any basis state in (24) is encoded into a product of two $n$-qubit GHZ
states, it is straightforward to show that for the encoded state (25), the
density operator of each qubit is given by $\frac 12\left( \left|
0\right\rangle \left\langle 0\right| +\left| 1\right\rangle \left\langle
1\right| \right) .$ This result means that the $n$-qubit quantum
information, originally carried by the $n$ ``message'' qubits, is hidden
over each qubit after encoding the state (24) into (25).

The encoding can be easily done by using Hadamard gates and CNOT gates. For
simplicity, we consider the case when each basis state in (24) is encoded
into a product of two $n$-qubit GHZ states both taking the same form. The
encoding operation is given by 
\begin{equation}
U_e=\prod\limits_{i=1}^{n-1}C_{n^{\prime }i^{^{\prime }}}\otimes
\prod\limits_{i=1}^{n-1}C_{ni}\otimes H_{n^{^{\prime }}}H_n\otimes
\prod\limits_{i=1}^nC_{ii^{^{\prime }}},
\end{equation}
where the $n$ ancillary qubits are initially in the state $\left|
00...0\right\rangle ;$ $H_n$ and $H_{n^{^{\prime }}}$ are Hadamard
transformation operations, respectively, acting on the ``message'' qubit $n$
and the ancillary qubit $n^{\prime }$; $C_{ii^{\prime }}$ is a CNOT
operation acting on the ``message'' qubit $i$ (control bit) and the
ancillary qubit $i^{\prime }$ (target bit);$C_{ni}$ is a CNOT operation
acting on the ``message'' qubit $n$ (control bit) and the ``message'' qubit $%
i$ (target bit); and $C_{n^{\prime }i^{\prime }}$ is a CNOT operation acting
on the ancillary qubit $n^{\prime }$ (control bit) and the ancillary qubit $%
i^{\prime }$ (target bit).

One possible application for hiding $n$-qubit quantum information over each
qubit is multi-qubit quantum information secret sharing among many receivers
in a network. As an example, let us consider this situation, i.e., Alice
needs to send $n$ qubits of quantum information to $2n$ receivers in a
network, but she wishes that each receiver cannot get any information
without other receivers' cooperation. To implement this, Alice can encode
the state (24) of her $n$ ``message'' qubits into the state (25) by using $n$
ancillary qubits, and then she sends one qubit of the $2n$ qubits to each
receiver through secure quantum channels. As shown above, since quantum
information is hidden over each qubit of the $2n$ qubits after the encoding,
it is clear that each receiver can not get any information from his/her
qubit, if no other receivers cooperate with him.

It should be mentioned that a general theory about quantum data hiding has
been proposed [35]. Althoug we treat a special case that a single party
cannot gain any information about the state, our main purpose is to wish to
present a concrete encoding scheme for hiding $n$-qubit information over
each qubit. This scheme also provides a good illustration of the
relationship between quantum data hiding and QECC already noted in [35, 36],
since it is straightforward to show that the above encoding is also
equivalent to a QECC correcting one erasure.

Taking into account the price which we will probably have to pay in
determining the error position, the fact that we have to know which qubit
goes ``bad'' (for example, if errors are accompanied by the emission of
quanta, they can in principle be detected) is a significant disadvantage of
erasure-error correction schemes over error correction schemes generally
working for unknown error positions. But again, it is compensated for by the
fact that we need a smaller number of ancillary qubits to construct a
quantum erasure-correcting code, for example, only one ancillary qubit is
required for one ``message'' qubit on average as far as the present code.
Also, as shown above, since the ``damaged'' particle is not involved in the
error recovery operations, the present code can still work in the case when
the interaction with environment leads to the leakage of a qubit out of the
qubit space.

As noted in [34], quantum erasure-correcting codes may be applied in $fault$ 
$tolerant$ $quantum$ $computing,$ which was proposed by Shor and permits one
to perform quantum computation and error correction with a network of
erroneous quantum gates [40]. Thus, the present code should be useful in a
small-scale $fault$ $tolerant$ quantum computing. Moreover, since quantum
information originally carried by the three ``message'' qubits is now hidden
over each physical qubit of the code, the present code may have some other
applications in quantum information processing and quantum communication,
such as quantum secret sharing [41] and quantum cryptography [42].

In conclusion, we have presented a six-qubit code for protecting three-qubit
quantum information against one erasure. The encoding, decoding and error
recovery operations, as shown here, are relatively straightforward. A
special feature of the error recovery method is that no extra ancillary
qubits and no measurement are required. The present code is also a
three-qubit quantum hidden information code over each qubit. In addition, we
have proposed an encoding scheme for hiding $multi$-qubit quantum
information over each qubit.

\begin{center}
{\bf ACKNOWLEDGMENTS}
\end{center}

This work was partially supported by US National Science Foundation
(EIA-0082499).


\begin{references}
\bibitem{s1}  E. M. Rains in $IEEE$ $Transactions$ $on$ $Information$ $Theory
$, Volume 45, Issue 1, January 1999, pp. 266-271.

\bibitem{s2}  P. W. Shor in $Proc.$ $35th$ $Annual$ $Symp.$ $\ on$ $%
Foundations$ $of$ $Computer$ $Science$ ( IEEE Computer Society Press, New
York 1994), pp. 124-134.

\bibitem{s3}  I. L. Chuang, R. Laflamme, P. W. Shor and W. H. Zurek, Science
270, 1633 (1995).

\bibitem{s4}  D. Deutsch, Proc. R. Soc. A 400, 97 (1985); ibid. 425, 73
(1989).

\bibitem{s5}  L. K. Grover, Phys. Rev. Lett. 79, 325 (1997).

\bibitem{s6}  P. W. Shor, Phys. Rev. A 52, R2493 (1995).

\bibitem{s7}  A. M. Steane, Phys. Rev. Lett. 77, 793 (1996).

\bibitem{s8}  R. Laflamme, C. Miquel, J. P. Paz, and W. H. Zurek, Phys. Rev.
Lett. 77, 198 \ (1996).

\bibitem{s9}  A. Ekert and C. Macchiavello, Phys. Rev. Lett. 77, 2585 (1996).

\bibitem{s10}  D. Gottesman, Phys. Rev. A 54, 1862 (1996).

\bibitem{s11}  C. H. Bennett, D. P. DiVincenzo, J. A. Smolin, and W. K.
Wootters, Phys. Rev. A 54, 3824 (1996).

\bibitem{s12}  A. M. Steane, Phys. Rev. A 54, 4741 (1996); A. M. Steane,
Proc. R. Soc. London A 452, 2551 (1996).

\bibitem{s13}  P. W. Shor, LANL eprint quant-ph/9605011; P. Shor and R.
Laflamme, Phys. Rev. Lett. 78, 1600 (1997).

\bibitem{s14}  D. P. DiVincenzo and P. W. Shor, Phys. Rev. Lett. 77, 3260
(1996).

\bibitem{s15}  W. H. Zurek and R. Laflamme, Phys. Rev. Lett. 77, 4683 (1996).

\bibitem{s16}  M. B. Plenio, V. Vedral, and P. L. Knight, Phys. Rev. A 55,
67 (1997).

\bibitem{s17}  A. R. Calderbank, E. M. Rains, P. W. Shor, and N. J. A.
Sloane, Phys. Rev. Lett. 78, 405 (1997).

\bibitem{s18}  E. Knill and R. Laflamme, Phys. Rev. A 55, 900 (1997).

\bibitem{s19}  D. W. Leung, M. A. Nielsen, Isaac L. Chuang, and Y. Yamamoto,
Phys. Rev. A 56, 2567 (1997).

\bibitem{s20}  J. Preskill, LANL e-print quant-ph/9705031.

\bibitem{s21}  C. H. Bennett and P. W. Shor, IEEE Trans. Inf. Theory 44,
2724 (1998).

\bibitem{s22}  L. Vaidman, L. Goldenberg, and S. Wiesner, Phys. Rev. A 54,
R1745 (1996).

\bibitem{s23}  L. M. Duan and G. C. Guo, Phys. Rev. A 57, 2399 (1998).

\bibitem{s24}  L. M. Duan and G. C. Guo, Phys. Rev. Lett. 79, 1953 (1997).

\bibitem{s25}  P. Zanardi and M. Rasetti, Phys. Rev. Lett. 79, 3306 (1997);
P. Zanardi, Phys. Rev. A 57, 3276 (1998).

\bibitem{s26}  D. A. Lidar, D. Bacon, and K. B. Whaley, Phys. Rev. Lett. 82,
4556 (1999); D. A. Lidar, I. L. Chuang, and K. B. Whaley, Phys. Rev. Lett.
81, 2594 (1998).

\bibitem{s27}  L. Viola and S. Lloyd, Phys. Rev. A 58, 2733 (1998); L.
Viola, E. Knill, and S. Lloyd, Phys. Rev. Lett. 82, 2417 (1999);

L. Viola, S. Lloyd, and E. Knill, Phys. Rev. Lett. 83, 4888 (1999).

\bibitem{s28}  D. Vitali and P. Tombesi, Phys. Rev. A 59, 4178 (1999).

\bibitem{s29}  G. S. Agarwal, Phys. Rev. A 61, 013809 (2000).

\bibitem{s30}  E. Knill, R. Laflamme, and L. Viola, Phys. Rev. Lett. 84,
2525 (2000); L. Viola, E. Knill and S. Lloyd, Phys. Rev. Lett. 85, 3520
(2000).

\bibitem{s31}  S. De Filippo, Phys. Rev. A 62, 052307 (2000).

\bibitem{s32}  P. Zanardi, Phys. Rev. A 63, 12301 (2001).

\bibitem{s33}  C. P. Yang and J. Gea-Banacloche, Phys. Rev. A 63, 022311
(2001).

\bibitem{s34}  M. Grassl, Th. Beth and T. Pellizzari, Phys. Rev. A 56, 33
(1997).

\bibitem{s35}  R. Cleve, D. Gottesman, and H. K. Lo, Phys. Rev. Lett. 83,
648 (1999); See also D. Gottesman, Phys. Rev. A 61, 042311 (2000).

\bibitem{s36}  N. J. Cerf and Richard Cleve, Phys. Rev. A 56, 1721 (1997).

\bibitem{s37}  T. Pellizzari, Phys. Rev. Lett. 79, 5242 (1997).

\bibitem{s38}  M. A. Nielsen and I. L. Chuang, $Quantum$ $Computation$ $and$ 
$Quantum$ $Information,$(Cambridge University Press, Cambridge, England,
2001).

\bibitem{s39}  W. D\"ur, G. Vidal, and J. I. Cirac, Phys. Rev. A 62, 062314
(2000).

\bibitem{s40}  P. W. Shor, in $Proceedings$ $of$ $the$ $37th$ $Symposium$ $%
on $ $Foundations$ $of$ $Computer$ $Science$ (IEEE, Los Alamitos, 1996), p
56.

\bibitem{s41}  M. Hillery, V. Buzek, and A. Berthiaume, Phys. Rev. A 59,
1829 (1999).

\bibitem{s42}  C. H. Bennett and G. Brassard, in $Proceedings$ $of$ IEEE $%
International$ $Conference$ $on$ $Computers$, $Systems$, $and$ $Signal$ $%
Processing$, $Bangalore$, $India$ (IEEE, New York, 1984); C. H. Bennett, G.
Brassard, and N. D. Mermin, Phys. Rev. Lett. 68, 557 (1992).
\end{references}
\end{document}